\documentclass{menu99}    
\usepackage{epsfig,psfig}

\begin{document}
    \setlength{\baselineskip}{2.6ex}

\title{Meson Production Experiments with Electromagnetic Beams using CLAS}
\author{S.A. Dytman \\
{\em Department of Physics and Astronomy, University of Pittsburgh,
Pittsburgh, PA  15260} }
\vspace{0.3cm}

\maketitle

\begin{abstract}
The CEBAF Large Acceptance Spectrometer (CLAS) started taking production
data in February, 1998.  It is capable of taking data for many reactions
simultaneously with moderate resolution.  Spectrometer properties and
running conditions are presented.  Data has been taken for a variety of
conditions with polarized and unpolarized beam and target.  Preliminary 
results of the initial experiments are shown and discussed.  
\end{abstract}

\setlength{\baselineskip}{2.6ex}

\section*{INTRODUCTION}

The CEBAF Large Acceptance Spectrometer (CLAS) is presently taking
data at a very large rate with essentially all of the initial complement
of detectors functioning at design specifications.  A primary CLAS goal 
of interest to this conference is the study of $N^*$
resonances for masses less than about 2.5 GeV.  Since these states
couple with various strengths to a large number of open inelastic
channels, a multi-faceted detector is required.  By covering 
about 80\% of $4\pi$ solid angle for single particles, many reactions can
be studied simultaneously.  The measured reactions 
have common systematic errors, correcting
a significant problem in previous $N^*$ analyses.  Presently accepted
$N^*$ properties are determined almost solely from experiments
using pion beams.  Thus, the new experiments will be sensitive
to $N^*$ states that couple weakly to $\pi N$.  A more complete description
of the apparatus can be found in the published talk of Bernhard 
Mecking~\cite{1}.  Another review of CLAS was given by Volker 
Burkert~\cite{2}.

The physics of $N^*$s with CLAS is the study of nonstrange and 
strange baryons, i.e. $N^*$, $\Delta$, $\Lambda$, and $\Sigma$ baryons.  
The first identification of $\Xi$ baryons in CLAS photoproduction events
has recently been reported.
These states decay to $\pi N$, $\eta N$, $\pi \pi N$, $\omega N$, 
$K \Lambda$, $K \Sigma$, and
many other final states.  Targets of liquid hydrogen, deuterium, ammonia 
(NH$_3$), $^3$He, 
$^4$He, carbon and iron have been run with real and 
virtual photons using 1.5-5.5 GeV electron beams.  By running at a
luminosity of $10^{34} cm^{-2}s^{-1}$, data on a large variety 
of reactions are being accumulated at an instantaneous event rate of 
about 2.5 kHz.  After the first year of
data taking, we have approximately equaled the number of 
events taken before CLAS for reactions accessible with traditional
small solid angle spectrometers and greatly exceeded it for more
complicated reactions. 

PDG lists about 40 $N^*$ states with a variety of
quantum numbers and varying degrees of certainty.  Presently, a
good qualitative understanding of {\it many} properties of these states using 
various versions of the Constituent Quark Model (CQM) exists.  However, 
this has not yet been linked to QCD
in any formal way and presently does not account for the meson cloud.
Although full QCD calculations on the lattice are believed to
include all effects correctly, efforts to date have been limited
to the quenched (no $g \rightarrow q \bar{q}$) approximation.  Recent 
improvements in numerical techniques and in computing power have been very 
impressive and significant results are expected.

Limitations in previous
data are very clear when extracting resonance properties.  The
main problem is to turn the experimental observables first into 
partial wave amplitudes (which carry the strength for each reaction in a
particular value of angular momentum and parity), then into the resonant
part of the amplitude.  Resonances are then found as poles or 
Breit-Wigner masses
and widths.  The latter step has more model dependence; a
review of that situation can be found in \cite{3}.  The new experiments hope 
to provide significant new information on two general
fronts- the spectra of states and their photocoupling amplitudes for
$\gamma N \rightarrow N^*$ as a function of $Q^2$.  A significant prediction
of the CQM is that many $N^*$ states are yet to be found, the so-called 
``missing
states''.  For example, the CQM predicts 22 nonstrange L=2 excitations, but
only about 10-12 have been identified.  There is an additional possibility 
that hybrid baryons
will have a sufficiently different electromagnetic response that they
will stand out from the normal baryons.  The incident photon causes a
transition from the nucleon ground state (proton and neutron) to the various
excited states.  The dominant excitation mechanism is often through the 
s channel (see Fig. 1).  

\parbox{8.5cm}{
\epsfig{figure=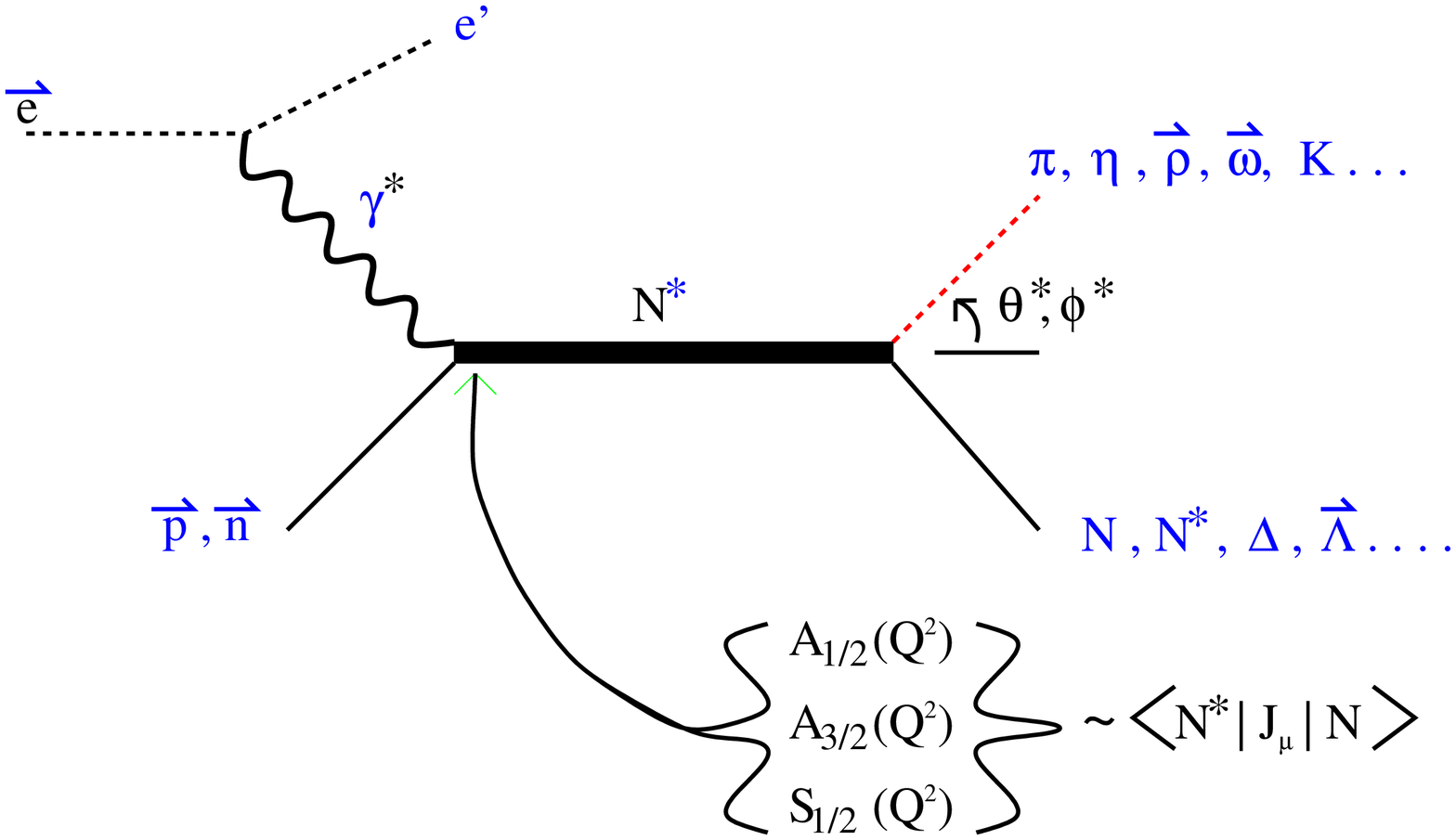,width=8cm,height=6cm}}
\parbox{7cm}{\vspace*{0.5cm}
\noindent
\parbox{5.2cm}
{\small \setlength{\baselineskip}{2.6ex} Fig.~1.  Excitation of a resonance 
via s channel.  Empirical studies show this to be the dominant reaction
mechanism for many final states, but u and t channel processes must be 
included for a
fully correct model.  Each event must be kinematically complete.  
The typical CLAS electroproduction experiment bins events in $Q^2$ 
and $W$ (virtual photon mass and invariant mass of the intermediate
state) and $\theta^*$ and $\phi^*$ (the decay angles of the meson
in the rest frame of the intermediate state.)}}
\vspace{0.2cm}

The transition strength for $\gamma N \rightarrow N^*$ can be determined in
various helicity states using a phenomenological analysis.  Since these 
strengths are defined as matrix
elements of the electromagnetic current acting between the $N$ and the $N^*$,
these results can be directly compared with any calculation of the
relevant wave functions.  Examples from a host of important issues
include trying to measure and explain the small size of the quadrupole 
excitation of
the $\Delta$ (P$_{33}$(1232)), i.e. the E2/M1 problem, and trying
to measure and explain the properties of the Roper (P$_{11}$(1440))
resonance.  A large body of data for the delta already exists; nevertheless,
CLAS will still greatly add to the electroproduction database.
For the Roper, the existing data is of poor quality and a 
significant improvement is possible in almost all reactions.  
Interpretation of the existing
data has been interesting because the CQM has trouble fitting the mass
and the photocoupling is poorly measured.
In addition, models suggesting interpretations as a hybrid
baryon or a meson-nucleon state rather than a predominantly
three quark state have been offered.

Some of the particles in Fig. 1 are displayed as vectors.  Polarized
beams at Jefferson lab are very common now and there is a
polarized target for protons and deuterons (ammonia).  A coherent 
Bremsstrahlung photon beam is being installed and is expected to
be tested next year followed by experiments emphasizing vector
meson production.
With the full coverage of the meson strong decay angles 
(e.g. $\phi \rightarrow K^+K^-$)
and the hyperon weak decay ($\Lambda \rightarrow p \pi^0$), some polarization 
information about the final state can be determined.  
Runs with polarized and unpolarized beams and targets have been taken.  
The events with unpolarized beam and target are the most fully analyzed 
and only that data will be presented here.

\section*{CLAS PROPERTIES}

CLAS was built by an international collaboration of physicists from about 30 
universities and national labs working with Jefferson lab personnel.  
Experiments are run by the CLAS collaboration.
CLAS has 6 almost identical sectors, each covering about 54$^\circ$ in $\phi$.
In Fig. 2, we show an
event with 2 tracks in opposite sectors.  The event has been classified
as $ep\rightarrow e'p\eta$ with the final state proton in the upper sector 
and the electron in the lower sector.  The detectors are labelled.

The detectors are conventional in design, but large on a nuclear
physics scale.  There are a total of 35,148 drift cells that
are used to track charged particles through the toroidal field.  
The drift chambers are divided into 34 layers of hexagonal cells.
Although the resolution of each cell is about 200 $\mu$m, overall system
resolution is presently about a factor of 2 larger.  Time of flight
resolution for the charged particles detected in the scintillators 
is about 140 ps for electrons.  The electromagnetic calorimeter has
energy resolution for photons and electrons of $\sigma_E/E \approx 0.1/
\sqrt{E}$.  The Cerenkov detector is run in threshold mode, so it
will fire only on electrons up to pion momentum of 2.8 GeV/c.  Empirical
studies have shown the Cerenkov to have efficiency larger than 99.5\%
in the area more than 10 cm from the edges.

\vspace{0.2in}
\parbox{8.5cm}{
\epsfig{figure=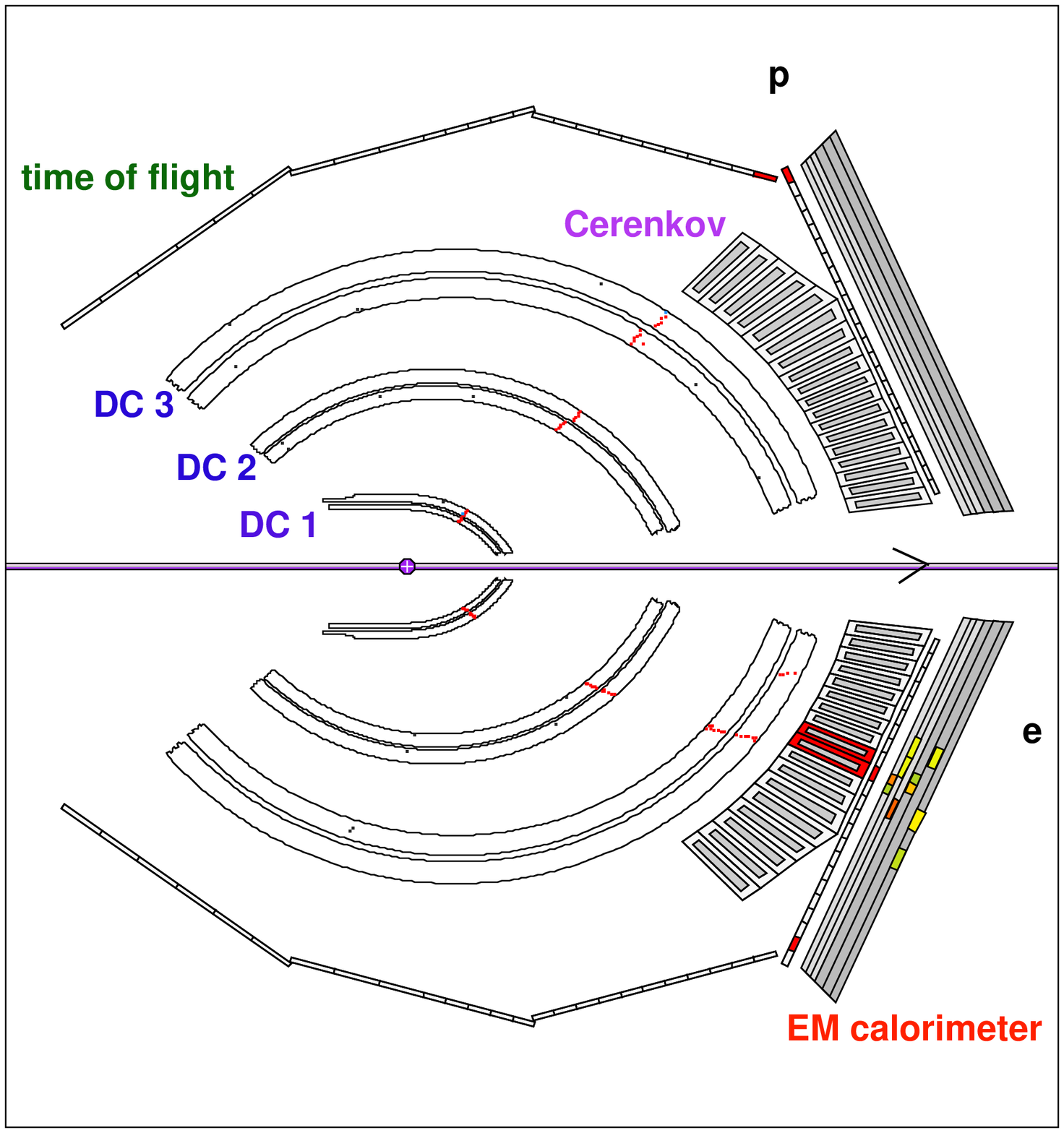,width=8cm,height=8cm}}
\parbox{7cm}{
\noindent
\parbox{5cm}
{\small \setlength{\baselineskip}{2.6ex} Fig.~2. Single event display
of an eta electroproduction event.  Various detector systems are labelled.
The arcs show locations of the 3 regions of drift chambers; the
dots represent drift cells where particles have deposited energy.  
The electron bends in the magnetic field toward the beam line, has large
energy deposition in the calorimeter, and fires
a Cerenkov counter.  The other track is a proton.  Particle identification
of charged hadrons is done via time of flight to the scintillators 
that cover the full CLAS solid angle.}}

In broad terms, polar angular ($\theta$) coverage for charged hadrons is 
about 8-140$^\circ$ with
moderate resolution.  For electrons and photons, the range is about 
20-47$^\circ$ due to the sizes of the Cerenkov detector and 
the electromagnetic calorimeter; in 2 sectors, shower counter coverage 
is extended to 75$^\circ$.  The momentum resolution is presently 
about 0.2\% in the forward direction and increases
to about 2-3\% for larger $\theta$.  Presently achievable resolution is
about a factor of 2 worse than design values.  Each event can have charged
particle tracks or neutral particles in any sector.  Events with 4 charged
particles have been completely reconstructed (see Table 2).  Electroproduction
cross sections must be binned in $Q^2, W, \theta^*,$ and $\phi^*$;
with the full running time allocated, this will provide an average of
a few hundred events in each of about a few hundred thousand bins
for the reactions with the largest cross sections.  Photoproduction
reactions have $Q^2$=0, no phi dependence, and in general better
statistics in each bin.

Technical papers covering all CLAS detectors are submitted and some are 
published~\cite{4}.

\section*{CLAS RUNNING EXPERIENCE}

With CLAS, many experiments take data simultaneously.
Thus, the normal nuclear physics delineation of experiments by final
states is not useful.  We label experiments by initial state and 
a wide range of final states are measured in one detector 
configuration.  For
example, the e1 run group covers all experiments with an electron
beam and an unpolarized
proton or neutron target.  That was the first run group to take beam 
and most of the results shown here are from that run.  The
beam can be photons or electrons, polarized or unpolarized; the target 
can be polarized or not.  To date, most runs used a liquid hydrogen
target- 5 cm long for e1 and 20 cm long for g1 
(unpolarized photon beam with an unpolarized hydrogen target).
Table 1 lists the run groups which have taken data as of August, 1999.
In each case, the trigger particle is listed in bold face.  For the
electron running, the trigger particle is the electron.  For photon
beam experiments, a charged particle in coincidence with a tagger
signal triggers each event.  The loose trigger is a key part of
fully utilizing large acceptance.  
Recent run cycles have produced a few billion events each.

\section*{FIRST LOOK AT RESULTS}

At this stage of analysis, a few reactions are well understood and reasonably
close to publication.  However, there are no final results of cross sections
presently available.  A sampling of distributions will be shown here.

The acceptance of CLAS is shown in Fig. 3 through the distribution
of $\pi^+$ seen in a small fraction of the first e1 run.  Traditional 
spectrometers detect particles over a few degrees in $\theta$ and $\phi$.  
Here, acceptance falls off at small values 
of $\theta$ due to support structures close to the beam line.  The 6 bands of 
missing events are due to the gaps between sectors filled by the magnet 
coils and detector supports.  
These fixed gaps are the major elements in the
acceptance calculation.  The detailed form of the acceptance 
depends on the value and sign of the magnetic field.  These
events were taken with 2.4 GeV beam, normal field (electrons bending
toward the beam line), and field value of 60\% of full strength (2250 A).

\begin{center}
\begin{table}
\caption{Summary of the first 1.5 years of CLAS data-taking.  Each run group
takes data for many reactions.  Polarized beam has become common at the
lab.  An arrow over a particle shows that it was significantly polarized.}
\vspace{0.2in}
\begin{tabular*}{6.0in}{@{\extracolsep{\fill}}|l|cl|}
\hline
                &run group  &reaction, trigger particle in bold\\
\hline
Feb-Mar, 98     &e1     &$e p \rightarrow {\bf e'} X$\\
May-June, 98    &g1     &$\gamma p \rightarrow {\bf c} X$ (c=charged hadron)\\
June, 98        &g6     &$\gamma p \rightarrow {\bf cc} X$\\
July, 98        &g1     &$\vec{\gamma} p \rightarrow {\bf c} X$\\
Aug-Dec, 98     &eg1    &$\vec{e} \vec{p} \rightarrow {\bf e'} X$\\
Jan-Apr, 99     &e1     &$\vec{e} p \rightarrow {\bf e'} X$\\
Apr-May, 99     &e2     &$\vec{e} A \rightarrow {\bf e'} X$  (A= $^{3,4}$He,$^{12}$C,Fe)\\
July, 99        &g6     &$\gamma p \rightarrow {\bf cc} X, \phi \rightarrow f_0 \gamma$\\
Aug, 99         &g2     &$\gamma d \rightarrow {\bf c} X$ \\

\hline
\end{tabular*}
\end{table}
\end{center}

\parbox{8.5cm}{
\epsfig{figure=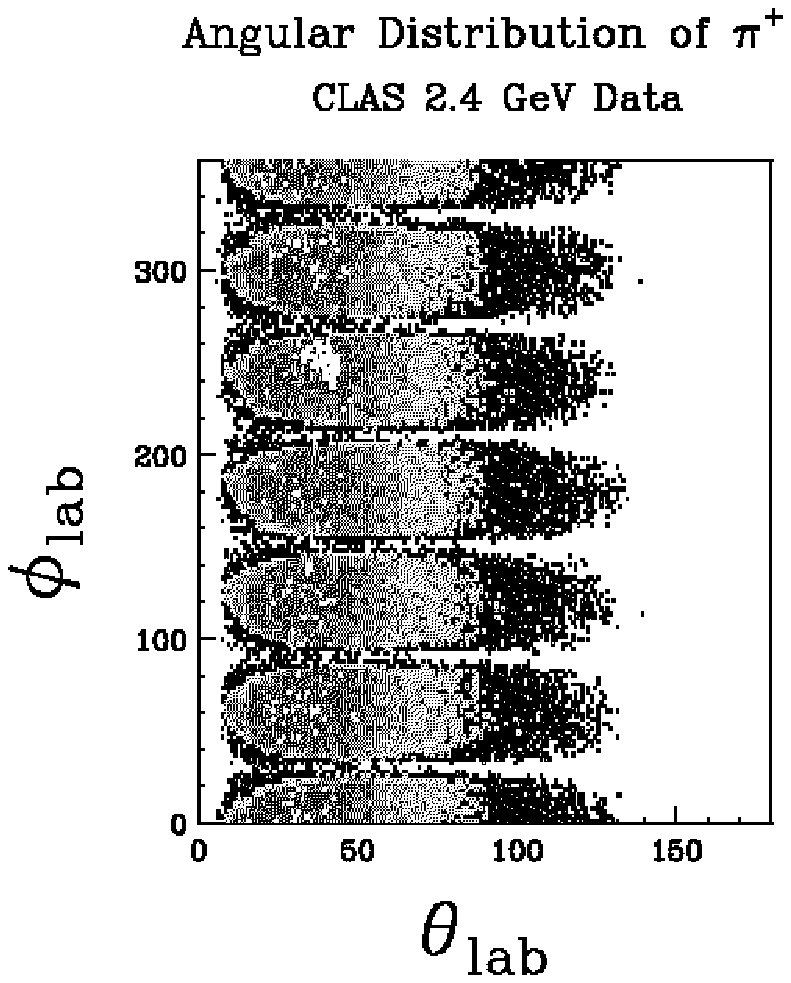,width=8cm,height=8cm}}
\parbox{7cm}{
\noindent
\parbox{5cm}
{\small \setlength{\baselineskip}{2.6ex} Fig.~3. Preliminary analysis of
the distribution of $\pi^+$ in the lab from part of the first e1 run.}}

Fig. 4 provides a broad, preliminary look at the physics to be measured with
CLAS.  We show $W$ distributions for 4 GeV beam energy and proton target.  
Final states were identified through missing mass
techniques.  The $Q^2$ of these events ranges between 1 and 3 
(GeV/c)$^2$.  Even though no
corrections have been applied and only a simple background subtraction
was used, the distributions are quite similar to the final cross sections.
The well-known 3 resonance regions are seen, peaked at $W \sim$ 1.2, 1.5, and
1.75 GeV.  However, the peaks are in different places and have different
strengths in the various reactions because each broad peak is a sum
of contributions from a few underlying states.  The reactions in the 
upper figures show strength in a variety of resonances; a detailed partial
wave analysis will be required to get strengths individual states .  
Both $\eta$ and $\omega$ distributions have a peak near threshold.
For $\eta$ electroproduction, the peak is known to be largely due to 
excitation of a single resonance, the S$_{11}$(1535).  These events have a 
largely isotropic angular distribution.  Since this is the first 
data for $\omega$ electroproduction
that wasn't dominated by diffractive processes, the peak is new.  It is 
probably due to
newly discovered coupling of known or new resonances to $\omega N$ since
the angular distributions have no hint of forward angle peaking close
to threshold.

\parbox{6.6cm}{
\noindent
\parbox{6cm}
{\small \setlength{\baselineskip}{2.6ex} Fig.~4. Preliminary analysis of
W distribution for various final states from part of the first e1 run.
Although no acceptance correction is applied, the acceptance in this variable
is fairly flat.  Sideband subtraction in the missing mass spectrum of a
neutral particle is sometimes used to isolate final states.}}
\parbox{8.5cm}{
\epsfig{figure=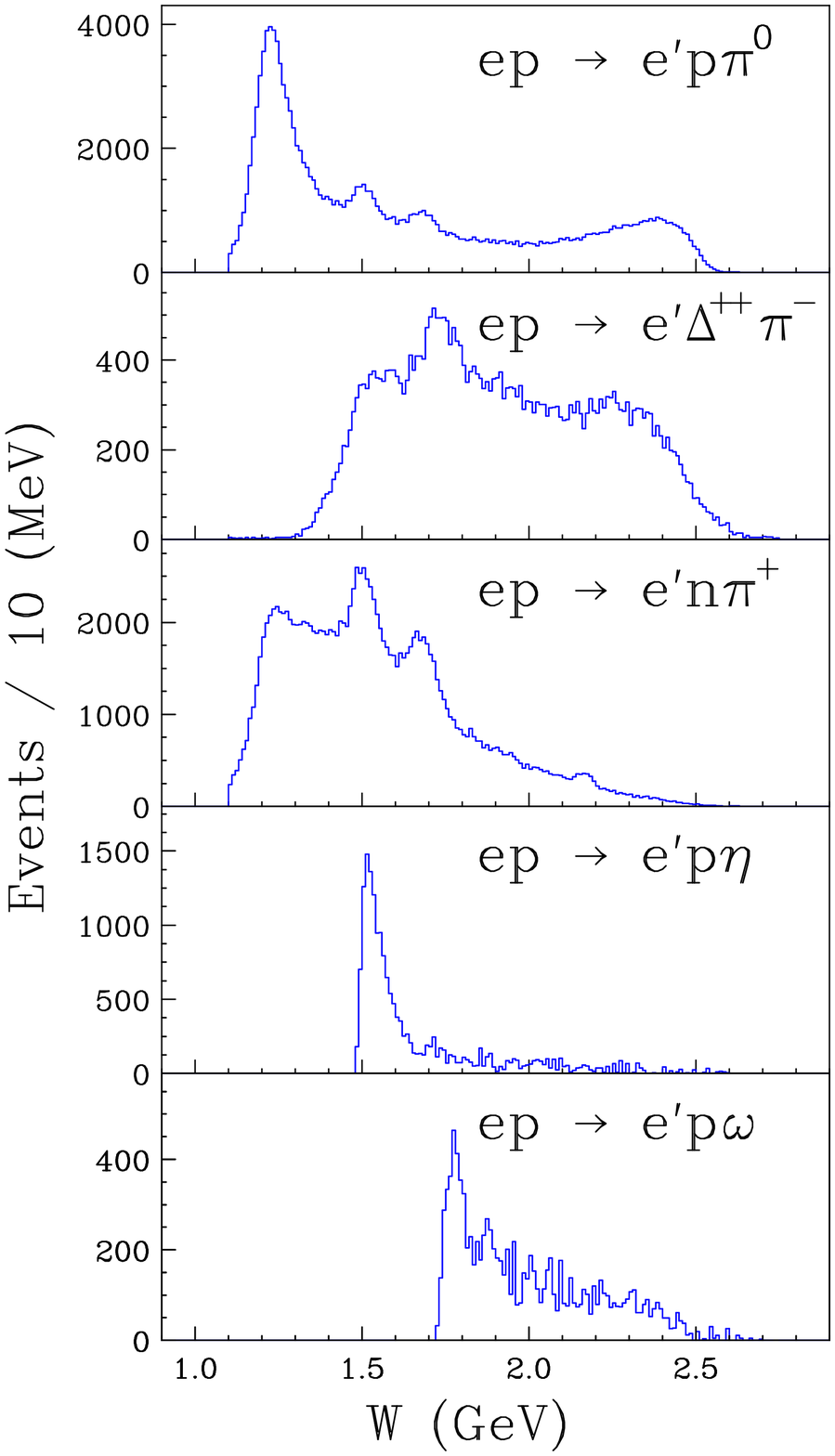,width=7cm,height=9cm}}

Table 2 lists the reactions that have been identified with CLAS
in the g1 and e1 data sets.  

\begin{center}
\begin{table}
\caption{Reactions identified in CLAS data as of October, 1999.  
The number of particles detected is also given.  Elastic
scattering and $\pi^+\pi^-$ production events have been very valuable
for calibration and efficiency measurements because they are overdetermined.}
\vspace{0.2in}
\begin{tabular}{|l|l||l|l|}
\hline
$e p \rightarrow e p $  & 1,2 & &  \\
$e p \rightarrow e' p \pi^0$  &2 &  $\gamma p \rightarrow p \pi^0$ & 1\\
$e p \rightarrow e' n \pi^+$ & 2 & $\gamma p \rightarrow n \pi^+$ & 1\\
$e p \rightarrow e' p \eta$  & 2 & $\gamma p \rightarrow p \eta$ & 1\\
$e p \rightarrow e' \Lambda K^+$ & 2 & $\gamma p \rightarrow \Lambda K^+$ & 1\\
$e p \rightarrow e' \Sigma K^+$ & 2 &  $\gamma p \rightarrow \Sigma K^+$ & 1\\
$e p \rightarrow e' p \omega$ & 3 & $\gamma p \rightarrow p \omega$ & 2\\
$e p \rightarrow e' p \phi$ & 3 & $\gamma p \rightarrow p \phi$ & 2\\
$e p \rightarrow e' p \pi^+ \pi^-$ & 3,4 & $\gamma p \rightarrow p \pi^+ \pi^-$ & 2,3\\
                           &   &  $\gamma p \rightarrow K^+ K^+ \Xi^- $ & 2 \\
$e p \rightarrow e' p \eta '$ & 4 &  $\gamma p \rightarrow p \eta '$ & 3\\
$e p \rightarrow e' p \eta \pi^+ \pi^-$ & 4 &  $\gamma p \rightarrow p \eta \pi^+ \pi^-$ & 2\\
$e p \rightarrow e' \Lambda (1520) K^+$ & 3 &  $\gamma p \rightarrow \Lambda (1520) K^+$ & 2\\
$e p \rightarrow e' \Lambda^0 K^+ \pi^-$ & 3 & $\gamma p \rightarrow \Lambda^0 K^+ \pi^-$ & 2\\
$e p \rightarrow e' \Lambda^0 K^{*0}$ & 3 & $\gamma p \rightarrow \Lambda^0 K^{*0}$ & 2\\
\hline
\end{tabular}
\end{table}
\end{center}

Acceptance calculations are in progress and largely understood.  The primary
contributor to acceptance is the geometric holes shown in Fig.~3.  These 
effects and detector inefficiencies are accounted for in angular 
distributions defined in Fig.~1 
($\theta^*$ and $\phi^*$) by Monte Carlo simulations.  Another important
effect in experiments with an electron beam is the radiative 
corrections.  An example of preliminary CLAS results, one of the 
$\phi^*$ distributions
for the $e p \rightarrow e' p \eta$ experiment, is shown in Fig.~5.  
Previous experiments had very limited $\phi^*$ coverage; still, the
angular distributions 
at $W \sim 1.5$ GeV over a large range in $Q^2$ were predominantly 
isotropic.  
(This means the interference response functions, $R_{LT}$ and 
$R_{TT}$ are small, but very poorly measured in previous experiments.)  
Thus, isotropic results within the
$\sim\pm$ 10\% errors of previous experiments can be expected.  The raw eta
yields (data points in the left figure) are decidedly nonisotropic, but the 
acceptance correction matches its shape to give a cross section that is 
isotropic.  The total correction factor is given as a line in the left plot.  
The radiative correction is about 15\% independent of angle.  For 
these points, the acceptance is about 35-40\%; the dips
come from situations when either the scattered
electron or the proton tend to be in the phi gaps.  A
quantity proportional to 
the cross section $d\sigma/d\Omega_{\eta^*}$ is shown in the right plot.

\parbox{9cm}{
\epsfig{figure=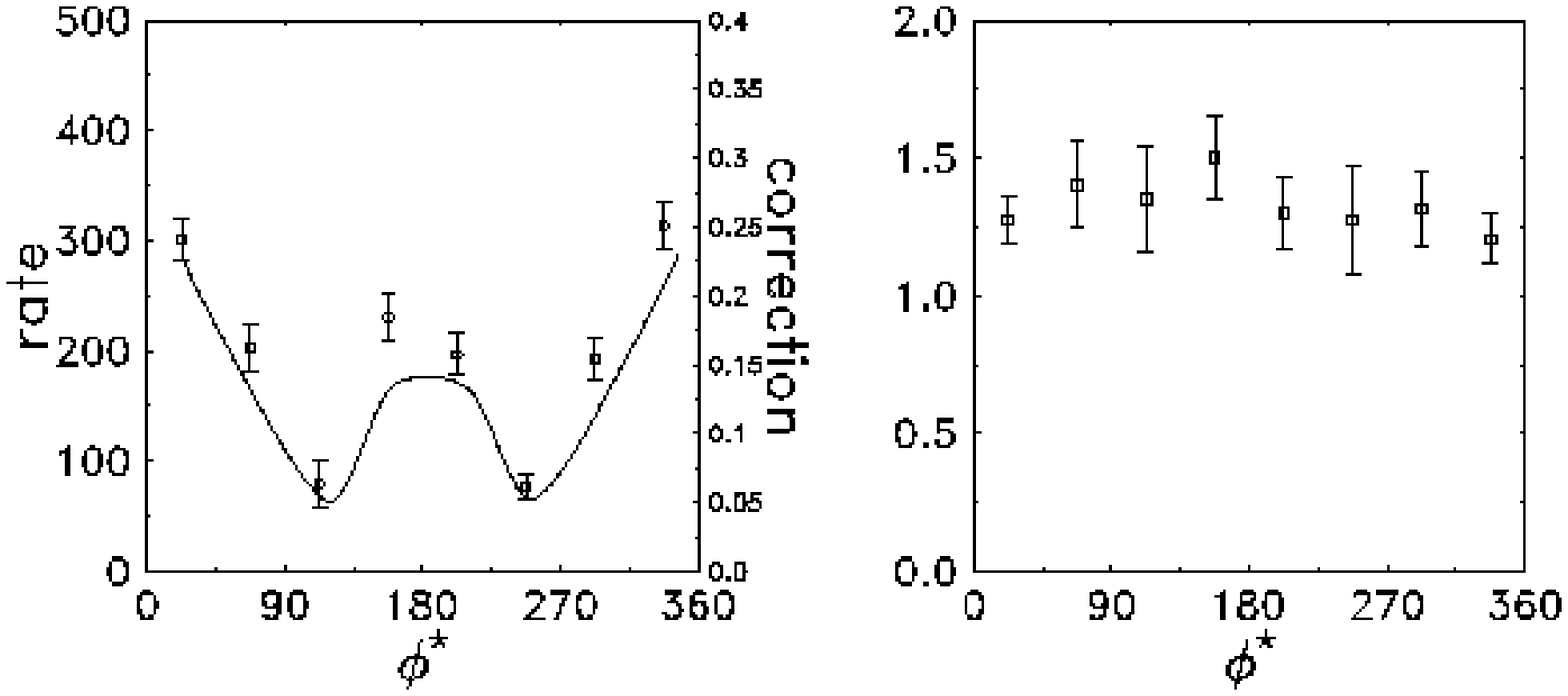,width=9cm,height=4.7cm}}
\parbox{7cm}{
\noindent
\parbox{5cm}
{\small \setlength{\baselineskip}{2.6ex} Fig.~5. $\phi^*$ dependence
for eta yield and
correction factor (left) and preliminary cross section
(right) for $e p \rightarrow e' p \eta$ at
$Q^2$= 0.75 (GeV/c)$^2$, $W$= 1.53 GeV, cos($\theta^*$)= -0.6.
Errors shown reflect statistical errors in the data and in
the Monte Carlo acceptance calculation.  Systematic errors for these
points are presently estimated as less than 10\%.}}

\section*{SUMMARY}
CLAS is now a working device.  In standard running mode, data is taken at an
instantaneous event rate of 2.5 kHz or about 0.5 Terabyte/day.  
It presently takes data for 10 months each year.  The collaboration 
has taken data with a variety of targets and with both photon and
electron beams.  A few run groups have recently 
accumulated a few billion events each in runs of about 2 months.  

At present, a significant number of reactions have been measured 
with statistical precision equal to or better than previous data.  
Some reactions are being explored for the first time.  Others are
being explored in kinematic regions not previously explored.
Systematic errors are presently under study.

Use of polarization has become common.  At the end of 1998, 2 months
were devoted to runs with polarized hydrogen target and polarized
electron beam.  Those experiments are important for isolating
specific spin parity intermediate states and for measuring an 
important contribution to the GDH Sum Rule at $Q^2 > 0$.
A diamond crystal and goniometer are being installed for production
of a transverse linearly polarized photon beam in 2000.
 
\section*{ACKNOWLEDGEMENTS}
The author is a member of the CLAS collaboration.  All results shown
have been taken and approved by the collaboration.  Special thanks for 
production of the figures used in this paper go to Stepan Stepanyan, 
Kui Young Kim, and Richard Thompson.

\bibliographystyle{unsrt}

\end{document}